\documentclass[pre,groupedaddress,twocolumn]{revtex4}
\usepackage[reqno,intlimits]{amsmath}
\usepackage{graphicx}
\usepackage{amssymb} 

\newcommand{\floatfig}[4][]{
  \begin{figure}[#1]
        #4
        \caption{#3}
        \label{#2}
  \end{figure}
}
\newcommand{\oneover}[1]{\frac{1}{#1}}

\newcommand{\e}{\ensuremath{\epsilon}}
\newcommand{\pea}{Parrondo \textit{et al}\ }
\newcommand{\mv}{\mathbf}

\newcommand{\floatplace}{htb}

\begin{document}
\bibliographystyle{apsrev}

\title[Winning combinations of history-dependent games]
      {Winning combinations of history-dependent games}

\author{Roland J. Kay}\email{roland.kay@physics.ox.ac.uk}
\author{Neil F. Johnson}
\affiliation{Physics Department, Oxford University, Parks Road, Oxford, OX1 3PU,
U.K.}

\begin{abstract}
The Parrondo effect describes the seemingly paradoxical situation in which two 
losing games can, when combined, become winning [Phys. Rev.  Lett. {\bf 85}, 24
(2000)]. Here we generalize this analysis to the case where both games are
history-dependent, i.e.  there is an intrinsic memory in the dynamics of each
game. New results are presented for the cases of both random and periodic
switching between the two games.
\end{abstract}

\date{\today}
\maketitle

\section{Introduction}


The Parrondo effect \cite{parrondo} is the counter-intuitive situation whereby
individually losing games somehow `cooperate' to produce a winning game. In
particular, these losing games can be combined {\em randomly} and yet the
effect still emerges. The intriguing aspect is that randomness in this system
is acting in a constructive way. Possible applications of this effect have been
suggested in several fields including biogenesis \cite{physics-and-life},
molecular transport \cite{kinderlehrer} \cite{heath}, random walks
\cite{random} and biological systems \cite{chaos}. Even in the social sciences,
`winning' models for investment have been reported \cite{th-appl-fin}.

Consider a gambling game in which the player has a time-dependent capital $X(t)$
where $t=0,1,2,\ldots,$ and whose evolution is determined by tossing biased
coins.  The rules as to which coins to toss, and hence the probability of
winning, are determined by the history, i.e. the game is history-dependent.  The
game can be divided into three regimes: winning, losing and fair (for which
$\langle X(t) \rangle$ is respectively an increasing, decreasing or constant
function of $t$).  Parrondo \textit{et al} \cite{parrondo} considered
combinations of such a history-dependent game B, as described above, and a
simple biased coin toss (i.e. game A which is history-{\em independent} and
hence has no memory).  In Parrondo \textit{et al}'s study, game
A is defined by the probability $p$ of $X(t)$ increasing, where $p=\oneover 2 -
\e$.  Hence game A is a losing game for $\e>0$.  Game B is defined by the
probabilities of four biased coins: $\{p_1,p_2,p_3,p_4\}$. The particular coin
played at a given time step depends upon the history of the game as shown in
Table \ref{tab:Bdef}.  Parrondo \textit{et al} showed that two losing games A
and B can be combined to yield a winning game, if the games are alternated
either periodically or at random.

\begin{table}[tb]
\begin{ruledtabular}
\begin{tabular}{c c c c c c}
Time step & Time step  & Coin    &\hspace{3mm}& Prob. of win & Prob. of loss \\
$t-2$     & $t-1$      & at $t$  &            & at $t$       & at $t$        \\
\hline
Loss & Loss & $B_1$ && $p_1$ & $1-p_1$ \\
Loss & Win  & $B_2$ && $p_2$ & $1-p_2$ \\
Win  & Loss & $B_3$ && $p_3$ & $1-p_3$ \\
Win  & Win  & $B_4$ && $p_4$ & $1-p_4$ \\
\end{tabular}
\end{ruledtabular}
\caption{Type B games consist of four coins. The coin to be played at time
step $t$ is determined by the results of the previous two time steps, as shown.}
\label{tab:Bdef}
\end{table}


The reason that Parrondo's paradox arises for combined A-B games is that losing
cycles in game B are effectively broken up by the memoryless behavior, or
`noise', of game A. The question therefore arises: what happens if \emph{both}
games are of type B, and hence have losing cycles? Can the losing cycle in one
game break up the losing cycle in the other in order to produce `winning
dynamics'? Since the answer is not obvious, and since the Parrondo effect
promises to have a variety of applications, it is important to establish whether
two history-dependent games will indeed produce a Parrondo effect. This provides
the motivation for the present study.

In this paper, we generalize the analysis of Ref. \cite{parrondo} to the case
where both games are history-dependent, i.e. there is an intrinsic memory in the
dynamics of each game. We find specific regimes which do indeed exhibit a
Parrondo effect. New results are presented for the cases of both random and
periodic switching between the two games.
The paper is organized as follows. In Sec. II we investigate random combinations
of two games of type B. In Sec. III we consider periodic combinations of such
games. In Sec. IV we investigate the effect of varying the switching
probability. Section V provides a summary.

\section{Random combinations of history-dependent games}

We now extend the analysis of Parrondo \textit{et al} to the case of two
history-dependent games of type B.
We define $\{p_i-\e\}$ and $\{q_i-\e\}$ as the probability sets defining the B
games and $\{r_i-\e\}$ as the probability set defining the combined game. We
follow \pea in only considering losing games which result by subtracting a small
quantity \e\ from each of the probabilities that define a fair game.
As in Ref. \cite{parrondo}, we can define a vector Markovian process $Y(t)$
based on the capital $X(t)$ as follows:
\begin{equation}
	\label{eqn:ydef}
	Y(t) = \begin{pmatrix}
	         X(t) - X(t-1) \\ X(t-1) - X(t-2) 
	       \end{pmatrix}
\end{equation}
$Y(t)$ can take four values $(\pm 1, \pm1)$. We label the four states of $Y(t)$
as shown in Table \ref{tab:states}.

\begin{table}[htb]
\begin{ruledtabular}
\begin{tabular}{c c c c c}
 \hfill & $Y(t)$    &\hspace{5mm} & State & \hfill\\
\hline
 & $(-1, -1)$ & & 1 & \\
 & $(+1, -1)$ & & 2 & \\
 & $(-1, +1)$ & & 3 & \\
 & $(+1, +1)$ & & 4 & \\
\end{tabular}
\end{ruledtabular}
\caption{Labels for the four possible states of the Markovian process $Y(t)$,
where $Y(t)$ is defined in terms of the capital $X(t)$ as prescribed by Eq.
\eqref{eqn:ydef}.}
\label{tab:states}
\end{table}

For $\e=0$ both B games must be fair. This is achieved by the condition
$(1-p_4)(1-p_3)-p_1p_2=0$ \cite{parrondo}. This yields the first two
conditions in Eq. \eqref{eqn:faircond}. For the combined game to be winning, we
obtain the final condition listed in Eq. \eqref{eqn:faircond}:
\begin{align}
	\label{eqn:faircond}
	(1-p_4)(1-p_3) & = p_1 p_2 \notag\\
	(1-q_4)(1-q_3) & = q_1 q_2 \\
	(1-r_4)(1-r_3) & < r_1 r_2 \notag \, .
\end{align}     
If the two B games are combined randomly, the probability set for the combined 
game is given by:
\begin{equation}
	\label{eqn:r}
	r_i = \alpha p_i + (1-\alpha)q_i
\end{equation}
where $\alpha$ is the probability that the game characterized by $\{p_i\}$ will
be chosen. We will typically take $\alpha = \oneover 2$.  The third condition in
Eq. \eqref{eqn:faircond} now becomes:
\begin{equation}
	\left(2-p_4-q_4\right) \left(2-p_3-q_3\right) < (p_1+q_1)(p_2+q_2)
	\, .
\end{equation}
Given that we require the initial games to be fair for $\e=0$, we can use the
first two conditions in Eq. \eqref{eqn:faircond} to substitute for $p_1$ and
$q_1$. Hence:
\begin{multline}
	\label{eqn:roughcond}
	\left(2-p_4-q_4\right) \left(2-p_3-q_3\right) < \\
	\left(\frac{(1-p_4)(1-p_3)}{p_2} +
	 \frac{(1-q_4)(1-q_3)}{q_2} \right) (p_2+q_2)
	 \, .
\end{multline}

\subsection{Special case $p_2=p_3,\, q_2=q_3$}

In order to reduce the number of free variables so that the different regions of
the parameter space can be displayed in a three dimensional figure, \pea
\cite{parrondo} made the restriction $p_2=p_3$.
Here we are going to reduce the number of free variables by appealing to the
first two conditions in Eq. \eqref{eqn:faircond}. These conditions give $p_1$
and $q_1$ in terms of \mbox{$\{p_j\},\, \{q_j\} \,\,\,\, (j=2,3,4)$} such that
both games are fair when $\e=0$.

We choose a particular game $\{p_i\}$ and then plot the regions in the parameter
space $(q_2,q_3,q_4)$ which enclose all games $\{q_i\}$ for which the Parrondo
effect is observed. 
Initially we treat the special case introduced by \pea using the parameter
space $(q_2=q_3,q_4)$, taking for the first B game:
\begin{equation}
	\label{eqn:pargameorig}
	\{p_i\} = \left\{\frac{9}{10}, 
	                 \frac{1}{4}, 
	                 \frac{1}{4}, 
	                 \frac{7}{10}
	          \right\}
\end{equation}
Rearranging Eq. \eqref{eqn:faircond} gives:
\begin{equation}
	\label{eqn:q1fair}
	q_1 = \frac{(1-q_4)(1-q_3)}{q_2} 
	\, .
\end{equation}
$q_1$ is a probability and is thus subject to the restriction $0<q_1<1$.
Therefore in order to be physically realized, the game $\{q_i\}$ must be
restricted as follows:
\begin{equation}
	\label{eqn:phys}
	q_4 > 1 + \frac{q_2}{q_3 - 1}
\end{equation}
and hence, in the special case where $q_3=q_2$:
\begin{equation}
	\label{eqn:physspec}
	q_4 > 1 + \frac{q_2}{q_2 - 1}
	\, .
\end{equation}
%

From Eq. \eqref{eqn:roughcond} the condition that defines the regions of
the parameter space in which two fair games combine to yield a winning game
is given by:
\begin{equation}
        \label{eqn:finalcondspec}
	q_4 \begin{cases}
		> 1 + \frac{(p_4 - 1)}{p_2}q_2   \qquad \text{if $q_2>p_2$}
		\\  \\
		< 1 + \frac{(p_4 - 1)}{p_2}q_2   \qquad \text{if $q_2<p_2$}\,.
	    \end{cases}
\end{equation}
Figure \ref{fig:speccond} depicts the regions of parameter space defined by 
Eqs. \eqref{eqn:finalcondspec} and \eqref{eqn:physspec} for $\{p_i\}$ given by
Eq. \eqref{eqn:pargameorig}.
In particular, Fig. \ref{fig:speccond} shows the region in which two fair games
combine to yield a winning game. This is equivalent to the region in which two
losing games combine to yield a winning game for some value of $\e >0$. 
In the appendix we derive an expression for the maximum value of \e\ for which
this remains true, $\e_{\rm{max}}$.

\floatfig[\floatplace]{fig:speccond}{
Parameter space for combination of two games in the special case
$p_2=p_3,q_2=q_3$ with $\{p_i\}$ given by Eq. \eqref{eqn:pargameorig}. Region I
depicts the area forbidden by Eq. \eqref{eqn:physspec}. Region II depicts the
area where the two games combine to yield a winning game. The white region in
between represents the area excluded by Eq. \eqref{eqn:finalcondspec}, in which
two fair games combine to yield a losing game.  The black dot represents the
parameters for \pea's original game as described in Ref. \cite{parrondo} for
which $\{q_1=\frac{1}{2},q_2=\frac{1}{2},q_3=\frac{1}{2},q_4=\frac{1}{2}\}$.
}{
\includegraphics[width=0.33\textwidth]{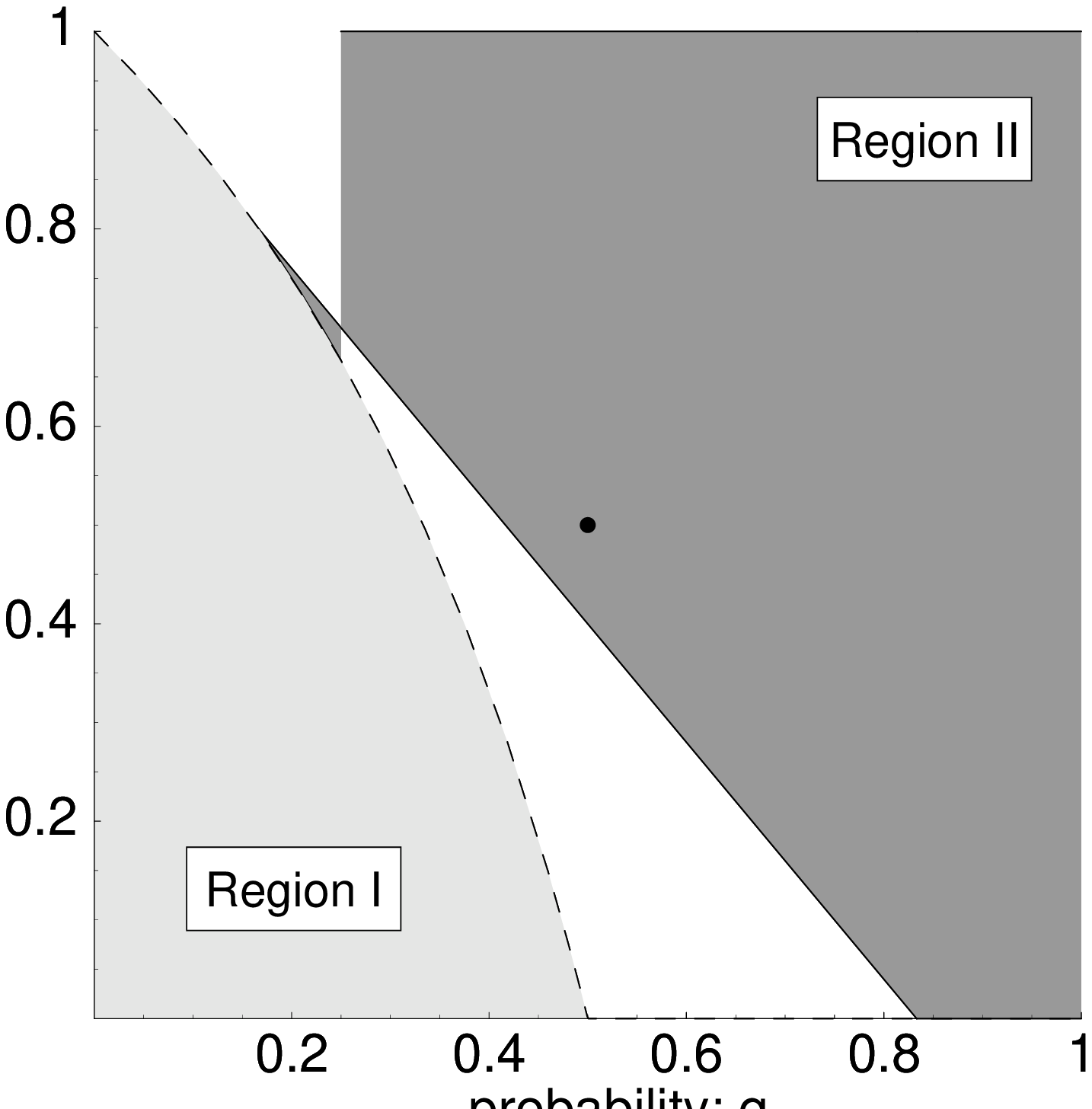}
}
\floatfig[\floatplace]{fig:contour}{
Plot of the maximum value of \e\ for which the two losing games $\{p_i\}$
defined by Eq. \eqref{eqn:pargameorig} and $\{q_1,q_2,q_2,q_4\}$, combine to
yield a winning game. The white lines indicate the locus of points where
$\e_{\rm{max}}=0$.  The regions in which $\e_{\rm{max}}<0$ correspond to two
winning games combining to yield a losing game. The region to the left of the 
dashed line is that excluded by Eq. \eqref{eqn:physspec}, as in Fig. 
\ref{fig:speccond}.
}{
\includegraphics[width=.43\textwidth]
{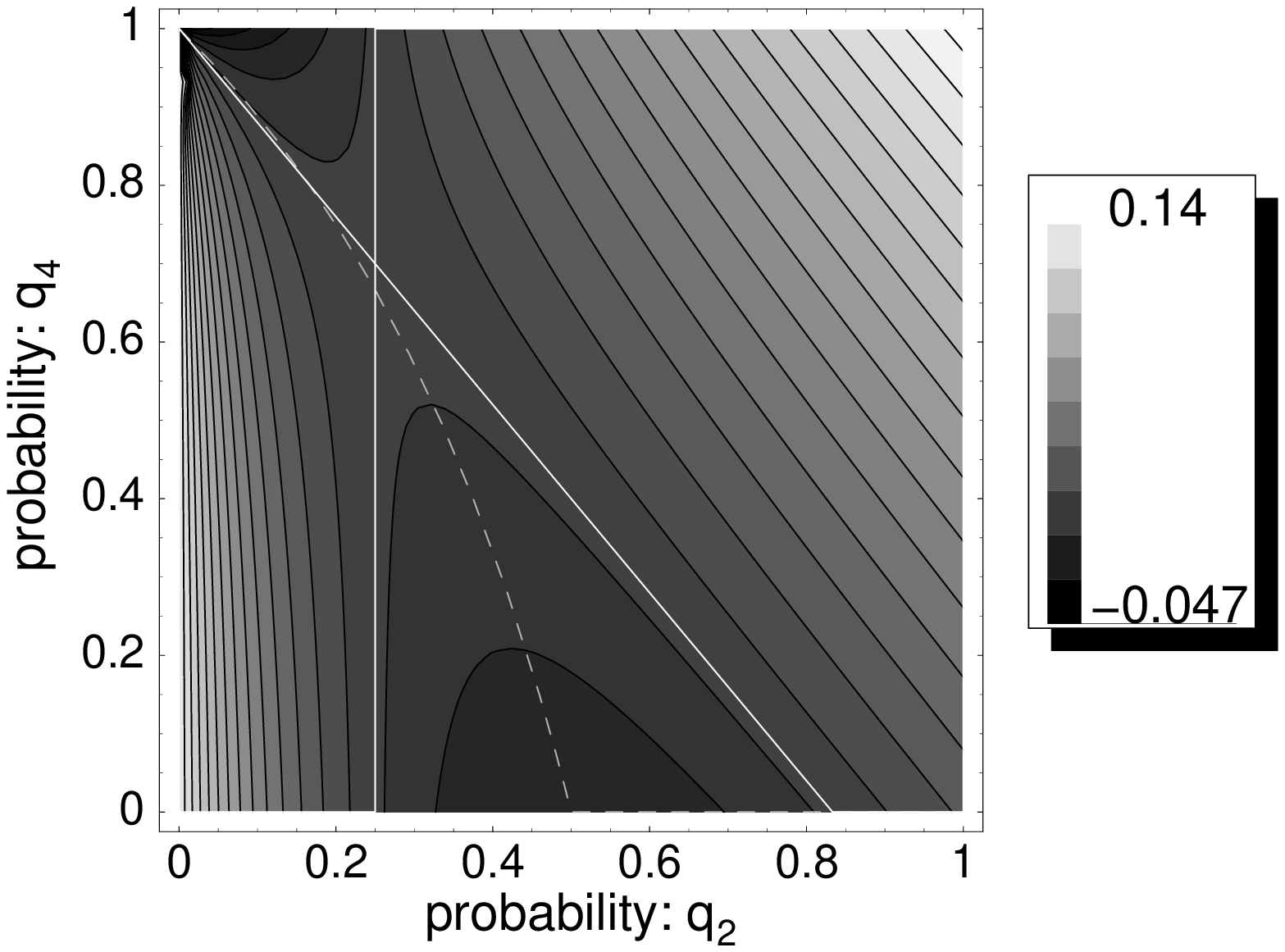}
}
Figure \ref{fig:contour} shows $\e_{\rm{max}}$, given by Eq. \eqref{eqn:epsilon_spec_fair} using the same game set $\{p_i\}$ as in Fig.
\ref{fig:speccond}. The value of $\e_{\rm{max}}$ is shown
for all possible games $\{q_i\}$ given $\{p_i\}$ defined by Eq.
\eqref{eqn:pargameorig}.  This plot demonstrates the robustness of the Parrondo
effect in the present case of two history-dependent games.

\subsection{General case $p_2\ne p_3,\, q_2\ne q_3$}
Now we drop Parrondo et al's restriction to $p_2= p_3$ and treat the general
case. 
From Eq. \eqref{eqn:roughcond}:
\begin{equation}
	\label{eqn:finalcondgen}
	q_4   \begin{cases}
		< 1 + \frac{(p_4 - 1)}{p_2}q_2 \qquad 
				\text{if $q_3<1+\frac{p_3-1}{p_2}q_2$}
		\\  \\
		> 1 + \frac{(p_4 - 1)}{p_2}q_2 \qquad 
				\text{if $q_3>1+\frac{p_3-1}{p_2}q_2$} \, .
	      \end{cases}
\end{equation}
Figure \ref{fig:gencond} depicts the regions of parameter space defined by Eq.
\eqref{eqn:finalcondgen} and Eq. \eqref{eqn:phys} for $\{p_i\}$ given by Eq.
\eqref{eqn:pargameorig}. Equation \eqref{eqn:finalcondgen} defines two regions
(labelled ``I'' and ``II'' in Fig. \ref{fig:gencond}). Equation \eqref{eqn:phys}
excludes almost all of region I in this case.
\floatfig[\floatplace]{fig:gencond}{
Regions of parameter space in which two fair games combine to yield a winning
game for $\{p_i\}$ given by Eq. \eqref{eqn:pargameorig}. The planes indicate the
boundaries of these regions (themselves marked ``I'' and ``II'').  The unmarked
regions are those in which the opposite effect occurs. The surface (and the
inset) show the boundary of the region forbidden by Eq. \eqref{eqn:phys}.  The
black dot represents \pea's original game as described in Ref. \cite{parrondo}
for which $\{q_1=\frac{1}{2},q_2=\frac{1}{2},q_3=\frac{1}{2},q_4=\frac{1}{2}\}$.
}{
\includegraphics[height=0.3\textwidth,width=0.4\textwidth]{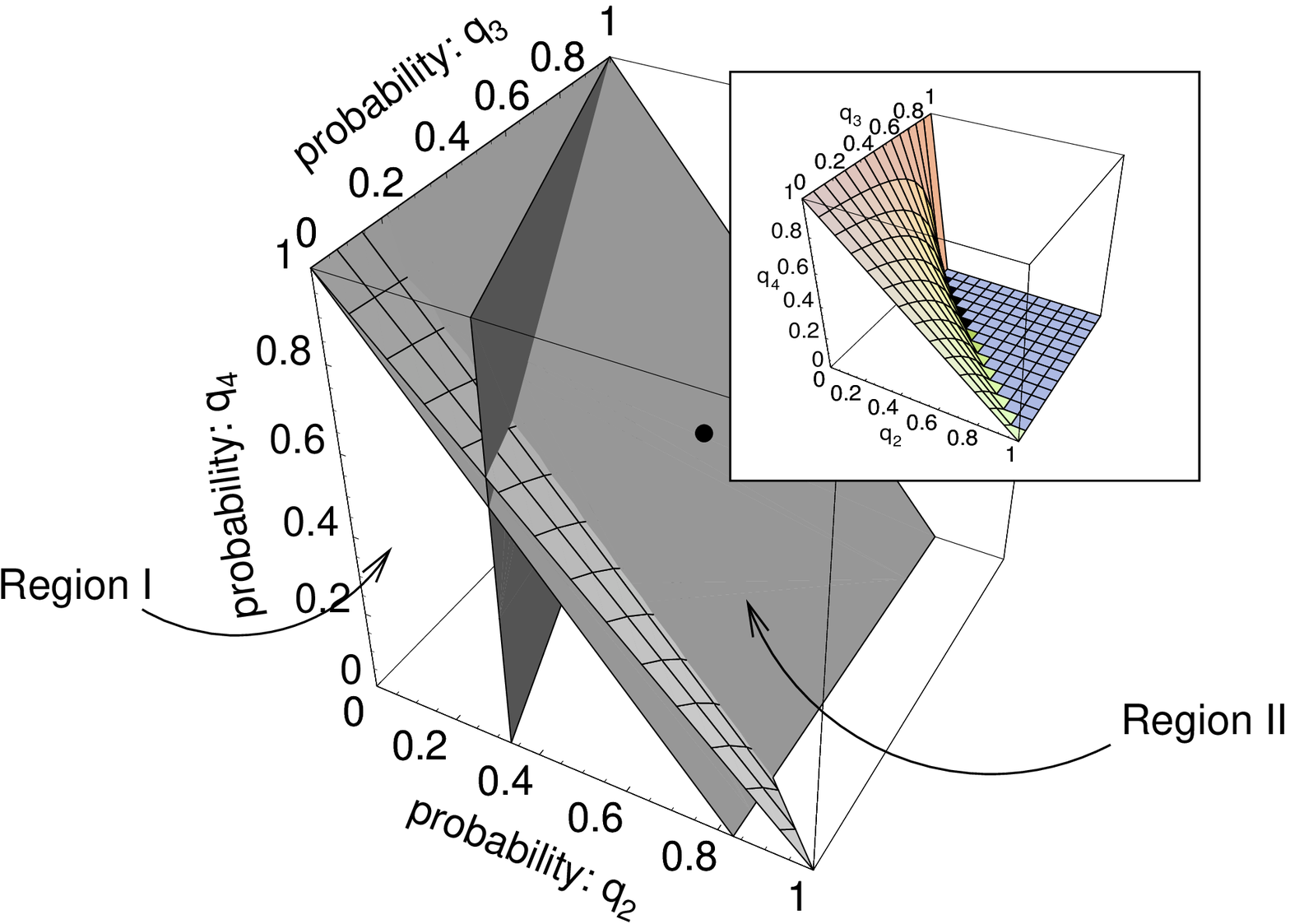}
}
%
%
%
An expression for $\e_{\rm{max}}$ in the general case is derived in the
appendix, Eq. \eqref{eqn:epsilon_gen_fair}, given that $\{q_i\}$ is a fair game
for $\e=0$.  In principle we could plot this over the 3D axes of Fig.
\ref{fig:gencond}. This would be the generalization of Fig. \ref{fig:contour}. 

The original combination of game A and game B due to \pea (see
Section I) represents a special case of our more general treatment.  The game
considered in Ref. \cite{parrondo} corresponds to combining $\{p_i\}$ as
defined by Eq. \eqref{eqn:pargameorig} with $\{q_1=\frac{1}{2},q_2=\frac{1}{2},
q_3=\frac{1}{2},q_4=\frac{1}{2}\}$.
In Figures \ref{fig:speccond} and \ref{fig:gencond} the black dot represents 
\pea's original game. In both cases it can be seen to lie in the region where
two losing games combine to yield a winning game.

Thus, we have derived expressions,  Eqs. \eqref{eqn:finalcondspec} and
\eqref{eqn:finalcondgen}, for the region of the parameter space in which the
Parrondo effect is observed to occur in the case of history-dependent games
being combined with equal probability ($\alpha=\frac{1}{2}$). We have also
derived expressions for the robustness of the effect, Eqs.
\eqref{eqn:epsilon_spec_fair} and \eqref{eqn:epsilon_gen_fair}.

\section{Periodic combinations of history-dependent games}

Next we investigate periodic combinations of games. Rather than randomly
selecting the game to be played at each time step, game $\{p_i\}$ is played $a$
times and then game $\{q_i\}$ is played $b$ times. This cycle is repeated
periodically.
Figure \ref{fig:fig2} shows the capital after 500 times steps, resulting from a
combination of two games for a range of values of $a$ and $b$.  The capital is
greater if the games are switched more frequently, as found by \pea for the
combination of a simple game A and a history-dependent game B.
\floatfig[\floatplace]{fig:fig2}{
Value of capital after 500 games averaged over an ensemble of $5\times 10^5$
runs. Games $\{p_i\}$ (defined by Eq. \eqref{eqn:pargameorig}) and $\{q_i\} =
\left\{\frac{2}{5}, \frac{3}{5}, \frac{3}{5}, \frac{2}{5}\right\}$ are combined
periodically for $\e=0$. Game $\{p_i\}$ is played $a$ times, then game $\{q_i\}$
is played $b$ times and so on.
}{
\includegraphics[height=.3\textwidth,width=.4\textwidth]{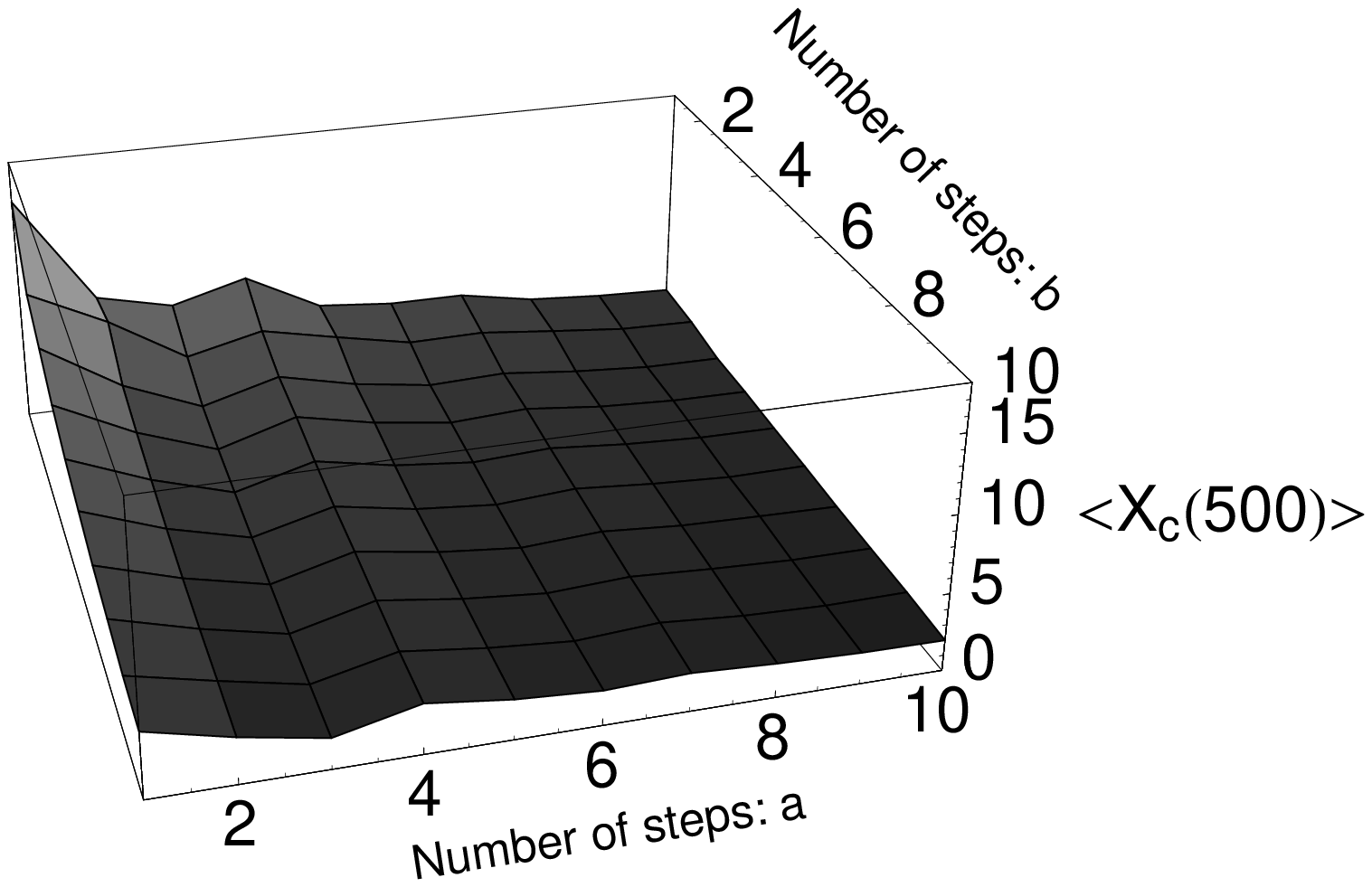}
}
%
%
The analysis for the periodic case is more complex than for the random case
because we can no longer appeal to a single game formed from a weighted
average of two games.

Let the elements of the vector $\mv{u}_i$, labelled $u_{i;j}$, be the
probability of the game being in state $j$ at time $t=i$.  The evolution of the
game from $\mv{u}_i$ to $\mv{u}_{i+a+b}$ can be described as follows:
\begin{gather}
	\label{eqn:perorig}
	\mv{u}_{i+a+b} = \mv{B}^b\mv{A}^a \mv{u}_i   \\
        \mv{A}=\begin{pmatrix}
		1-p_1 & 0     & 1-p_3 & 0     \\
		p_1   & 0     & p_3   & 0     \\
		0     & 1-p_2 & 0     & 1-p_4 \\
		0     & p_2   & 0     & p_4   \\
	\end{pmatrix}  \label{eqn:A} \\
	\mv{B}=\begin{pmatrix}
		1-q_1 & 0     & 1-q_3 & 0     \\
		q_1   & 0     & q_3   & 0     \\
		0     & 1-q_2 & 0     & 1-q_4 \\
		0     & q_2   & 0     & q_4   \\
	\end{pmatrix} \label{eqn:B}
\end{gather}
Clearly this is not a homogeneous Markovian process because the transition
matrix is not time-independent. 

In order to proceed we define a homogeneous Markovian process described by the
transition matrix $\mv{T}_0= \mv{B}^b\mv{A}^a$ with time steps 
$t=(a+b)i$, where $i=0,1,2,3,...$.
Consider a large ensemble of games described by Eq. \eqref{eqn:perorig} in the
long time limit. Select one of these games at random.  The stationary state
$\mv{S}_0$ of
the homogeneous game defined by $\mv{T}_0$ gives the probability that
the selected game will be in each of the four possible states  (see Table
\ref{tab:states}) at times $t=(a+b)i$, where $i=0,1,2,3...$.
This stationary state is given by the solution to the equation 
$\mv{S}_0=\mv{T}_0\mv{S}_0$.
Now we define a new homogeneous process, $\mv{T}_1$, by cyclically
permuting the matrices in $\mv{T}_0$ once to the right (e.g. if
$\mv{T}_0=\mv{BAA}$, then $\mv{T}_1 = \mv{ABA}$). The stationary state of this
process gives the probabilities of finding the game, selected from the ensemble,
in each of the four possible states at times $t=(a+b)i+1,$ where $i=0,1,2,3...$ 

\floatfig[\floatplace]{fig:explanation}{
Illustration of the time steps at which the stationary states $\mv{S}_n$ of the
transition matrices $\mv{T}_n$ give the probability of finding the combined game
in each of the possible states. $N$ is an integer and we take the long time
limit (i.e.  $N\rightarrow\infty$).
}{
\includegraphics[width=.48\textwidth]
{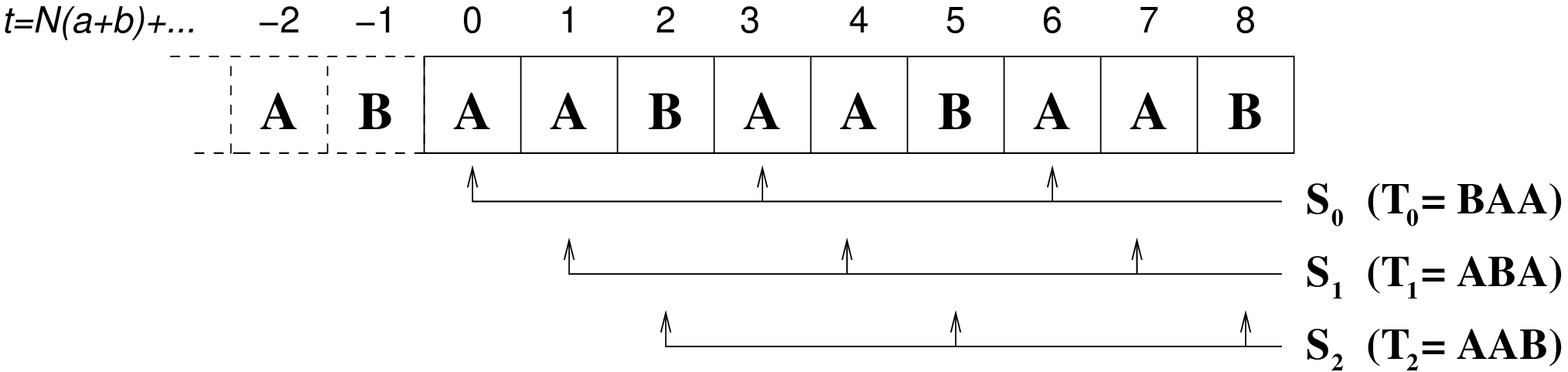}
}

In the general case the game formed from the $n$'th cyclic permutation of
$\mv{T}_0$, $\mv{T}_n$, gives us the probability of finding the selected game in
each state at times $t=(a+b)i+n,$ where $i=0,1,2,3...$. This is illustrated by
Fig. \ref{fig:explanation}.
We can calculate the overall probability of a win at time steps $t=(a+b)i+n$ by
taking the dot product of the stationary state of the transition matrix
$\mv{T}_n$, with a vector formed from the probabilities of each of
the coins  from the game played at that time step.
These vectors are $\mv{p}=(p_1\,\,p_2\,\,p_3\,\,p_4)$ and
$\mv{q}=(q_1\,\,q_2\,\,q_3\,\,q_4)$, where $\mv{p}$ corresponds to $\mv{A}$
in Eq. \eqref{eqn:A} and $\mv{q}$ corresponds to $\mv{B}$ in Eq. \eqref{eqn:B}.
The matrix to the right of the product in $\mv{T}_n$ corresponds to the game
that will be played at time step $t=(a+b)i+n$. Therefore, if the matrix to the
right is $\mv{A}$ we must take the dot product with $\mv{p}$. If it is
$\mv{B}$, we must take the dot product with $\mv{q}$.

An expression for the average probability $P_\text{win}$ of a win for the
combined game, can thus be found by averaging over all possible cyclic
permutations of $\mv{T}_0$. The gradient, $\text{grad}[\langle X_c(t)\rangle]$,
is then given by Eq. \eqref{eqn:pregrad}, as before.
The resulting expressions are lengthy. Each set of values of $a$ and $b$ yields
an expression for $\text{grad}[\langle X_c(t)\rangle]$ in terms of
$\{p_i\},\{q_i\}$, where $i=1,2,3,4$. These expressions are too complicated to
set out here explicitly. However, we can numerically plot the analytic
equivalent of Figure \ref{fig:fig2}: this is what we have essentially done in
Figure \ref{fig:theoryfig2}. The lines show the analytic prediction for the
average capital after 500 time steps, $\langle X_c (500) \rangle$, found by
multiplying Eq. \eqref{eqn:pregrad} by 500. Each line corresponds to a slice
through the surface in Fig. \ref{fig:fig2} at constant $b$.  The error bars
indicate one standard deviation on the mean over ten ensemble averages of the
numerical game.  Each ensemble average comprises an average over 50,000
individual runs.

We can see that the numerical and analytic results agree to within one standard
deviation. This confirms that the equations generated by the analysis presented
in this section are indeed correct.
Thus, we have derived expressions for the robustness of the Parrondo effect when
two history-dependent games are combined periodically.

\floatfig[\floatplace]{fig:theoryfig2}{
Comparison of analytic and numerical results for periodic switching. The lines
show the analytic predictions at constant $b$, the number of steps of game
$\{q_i\}$, for the capital after 500 time steps. The error bars show one
standard deviation on the mean over 10 separate ensemble averages, each
comprising 50,000 numerical runs.
}{
\includegraphics[height=.3\textwidth,width=.43\textwidth]{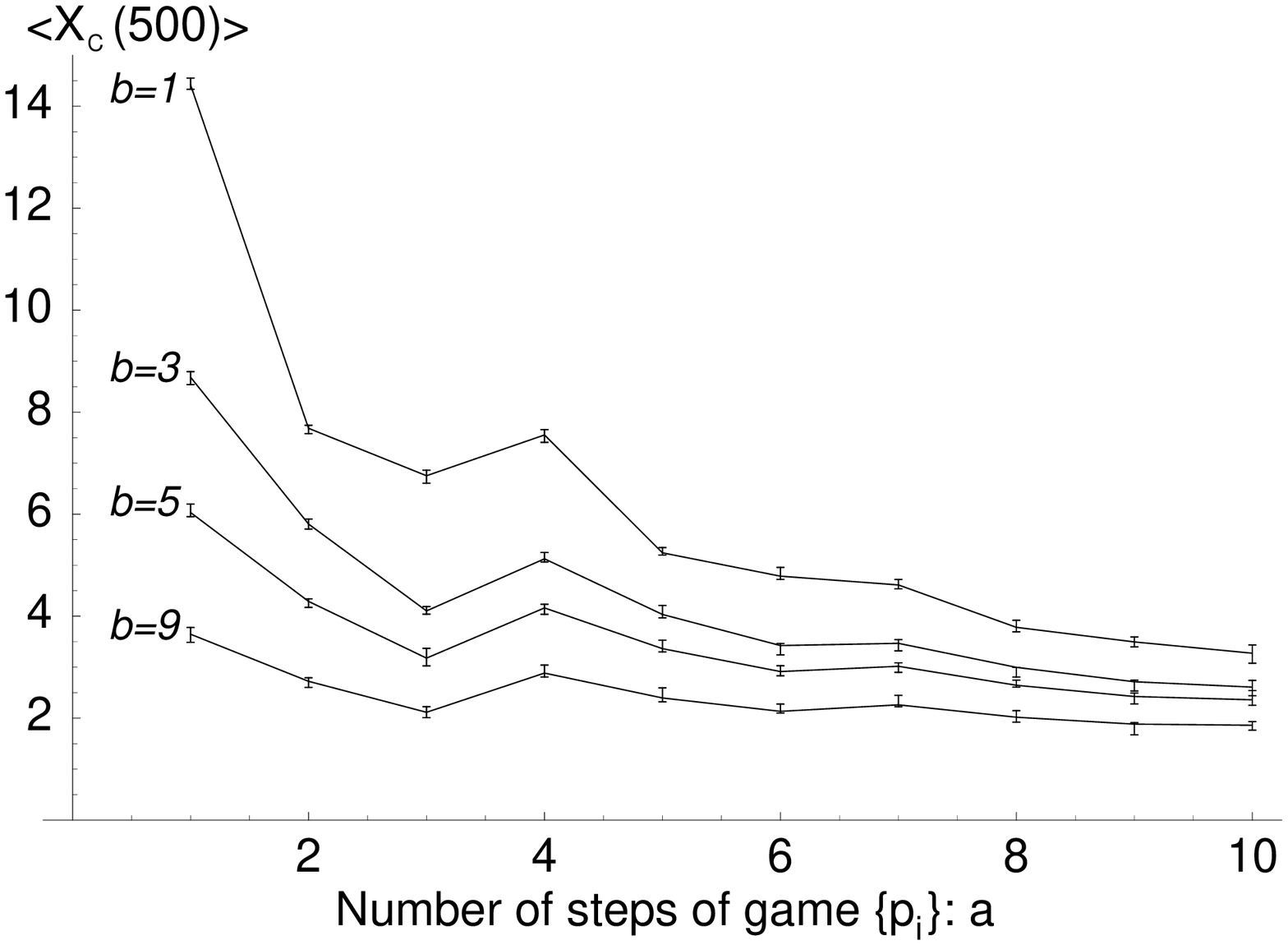}
}

\section{Varying the switching probability in the random case}

We now examine the dependence of the capital on the switching probability in the
case that the games are randomly combined. Figure \ref{fig:fig2rand} shows the
capital after 500 iterations plotted against the probability per iteration
$\alpha$ that the game $\{p_i\}$ will be chosen.  The curve is symmetric and
demonstrates that the capital is greatest if the games are switched with equal
probability.
When implementing the games it is necessary to assign values to the results of
the coin tosses at times $t=-2,-1$ in order to seed the game.  This arbitrary
choice introduces transients which can slightly bias the final results. However,
by allowing the game to first run for 100 iterations, this effect can be
eliminated.

The curve in Fig. \ref{fig:fig2rand} represents the capital predicted by Eq.
\eqref{eqn:grad} plotted for all $\alpha$ with the same $\{p_i\}$ and $\{q_i\}$.
The error bars show one standard deviation either side of the mean capital,
averaged over an ensemble of 10 runs.  The agreement between the theoretical
curve and numerical data is therefore better than one standard deviation.
For $\alpha=0 \text{ or } 1$ we find $\langle X_c(500) \rangle = 0$. These
values correspond to just playing one B game or the other. Since both games are
fair for $\e=0$ the average capital is zero. The fact that the curve is then
positive for all values of $\alpha$ means that combining the games with any
probability $0<\alpha<1$ leads to a winning combined game.

\floatfig[\floatplace]{fig:fig2rand}{
The value of the capital after 500 games averaged over an ensemble of 500,000
runs.  The same games as in Fig. \ref{fig:fig2} are combined, but this time
randomly.  Game $\{p_i\}$ is chosen with probability $\alpha$.  The solid line
shows the theoretical result predicted by Eq. \eqref{eqn:grad} while the error
bars (which are barely visible since they are so small) indicate one standard
deviation about the mean capital over an ensemble of 10 runs.
}{
\includegraphics[height=.3\textwidth,width=.43\textwidth]{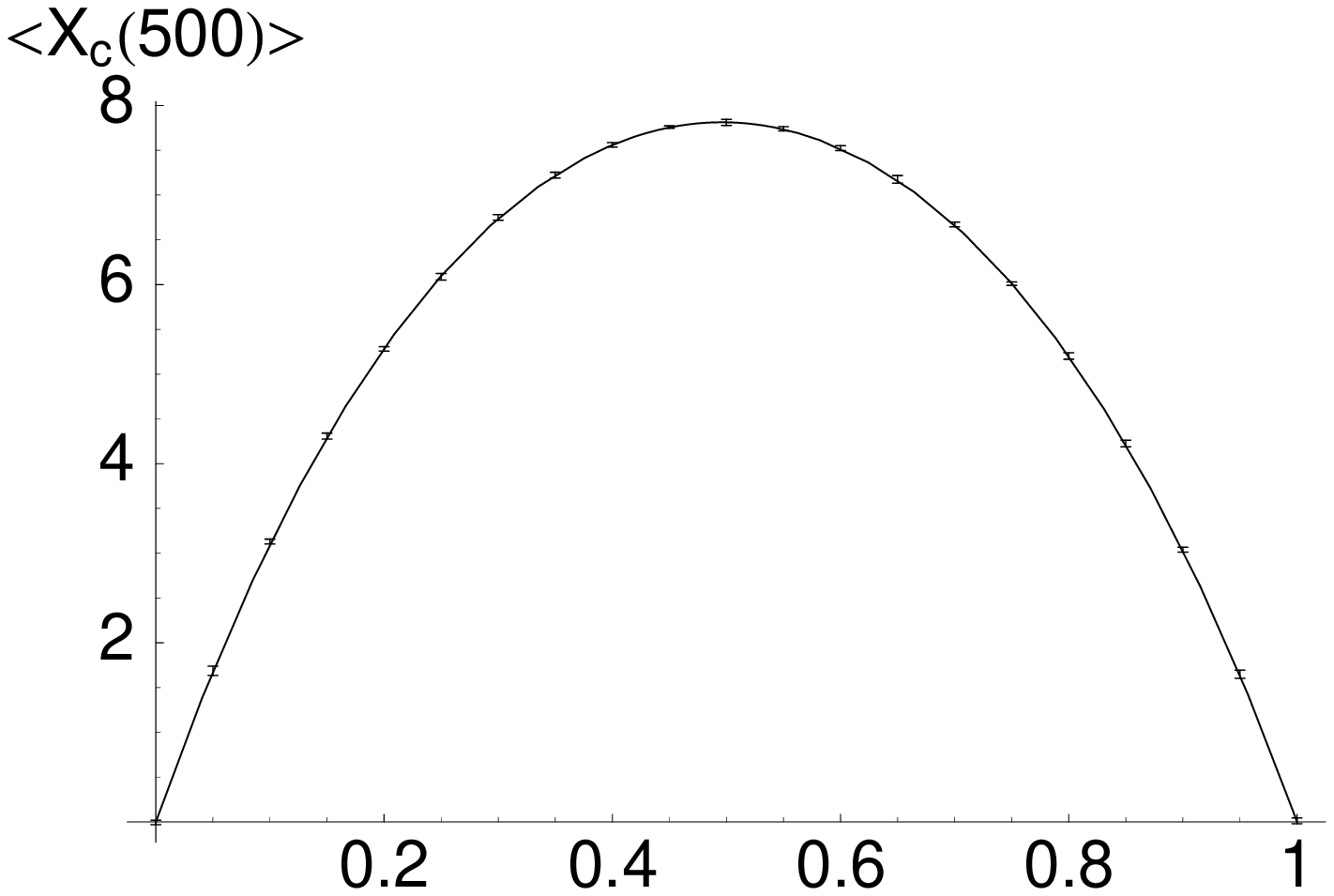}
}

\pea showed that combining two losing games could lead to a winning game because
switching between the games can break the cycles which cause the ``bad'' coins
to be over-played \cite{parrondo}. We might therefore expect that 
switching between the games frequently (either by reducing $a$ and $b$ or by
choosing a switching probability close to $0.5$) results in the largest capital
gain since this will break up the cycles most effectively. This is indeed 
confirmed by the results in Fig. \ref{fig:fig2} and \ref{fig:fig2rand}.


\section{Summary}

We have demonstrated that the apparently paradoxical effect of two losing games
combining to produce a winning game also applies to combinations of two 
history-dependent games. 
We derived expressions for the regions of the parameter space in which the 
effect is observed for both random and periodic combinations of these games.
%
We derived expressions for the gradient of the average capital and hence the
robustness of the Parrondo effect for games combined randomly
or periodically. 

Our work has therefore expanded the understanding of the Parrondo effect by
demonstrating its existence for new combinations of history-dependent games.  We
are now faced with the more general question as to what property of the
constituent games guarantees that the Parrondo effect will be observed.  In
addition if we were to combine many games of a more general nature than those
used to date, how could we predict whether the effect would emerge? 
We hope that the present work will stimulate further research on such questions,
in addition to pursuing applications of the Parrondo effect itself.

\begin{acknowledgments}
We thank Adrian Flitney for helpful comments and suggestions.
\end{acknowledgments}

\appendix
\section{Derivation of \NoCaseChange{ {\large $\e_{\rm{max}}$} } for random combinations of games }
In this appendix we derive expressions for the maximum value of $\e$, 
$\e_{\rm{max}}$, 
for which two games which are fair for $\e=0$ combine to yield
a winning game.

Let $P_\text{win}$ and $P_\text{lose}$ be the probabilities, per iteration for
the combined game, that $X_c(t)$ will increase or decrease respectively. Then
the gradient of the capital line for the combined game ${\rm grad}[\langle
X_c(t) \rangle]$ is given by:
\begin{align}
\label{eqn:pregrad}
	{\rm grad}[\langle X_c(t) \rangle] 
			&= 2P_\text{win} -1	\, .
\end{align}
One can derive the following expression for $P_\text{win}$ for a
game $\{r_i\}$:
\begin{equation}
	P_\text{win} = \frac{r_1(r_2+1-r_4)}{(1-r_4)(2r_1+1-r_3)+r_1r_2} \, .
\end{equation}
Substituting for $\{r_i\}$ from Eq. \eqref{eqn:r} 
yields:
\begin{multline}
\label{eqn:grad}
{\rm grad}[\langle X_c(t) \rangle] =  \\
\frac{2 \times\ldots}
     {[(1-\alpha)q_1+\alpha p_1] [(1-\alpha)q_2 + \alpha p_2] +\ldots}
     \\
\frac{\ldots [(1-\alpha)q_1 + \alpha p_1] \times\ldots}
     {\ldots (1 - 2(\alpha - 1)q_1+(\alpha-1)q_3+2\alpha p_1 - \alpha p_3)
     \times\ldots} 
     \\
\frac{\ldots [1 + (1-\alpha)q_2 - (1-\alpha)q_4  + \alpha(p_2 - p_4)]}
     {\ldots [1+(\alpha-1)q_4- \alpha p_4]}
     -1
\end{multline}
Now we reintroduce \e\ via the transformations $p_i\rightarrow p_i-\e$,
$q_i\rightarrow q_i-\e$ to obtain:
\begin{multline}
\label{eqn:gradepsilon}
{\rm grad}[\langle X_c(t) \rangle] =  \\
\frac{2 \times \ldots}
     {[\alpha p_1 +(1-\alpha)q_1-\e] [\alpha p_2 +(1-\alpha)q_2-\e] +\ldots}
     \\
\frac{\ldots [\alpha p_1 + (1-\alpha)q_1-\e] \times \ldots}
     {\ldots [-2\alpha p_1 +\alpha p_3 -2(1-\alpha)q_1+(1-\alpha)q_3+\e-1]
     \times\ldots} 
     \\
\frac{\ldots [\alpha(p_2-p_4)+(1-\alpha)(q_2-q_4)+1]}
     {\ldots [\alpha p_4+(1-\alpha)q_4 -\e -1]}
     -1
\end{multline}
Any games $\{p_i\},\{q_i\}$ defined as above will be losing games for all values
of $\e >0$. Thus in order to find the maximum value of \e\ for which two losing
games combine to yield a winning game, we must find the value of \e\ for which
${\rm grad}[\langle X_c(t) \rangle] = 0$. We shall consider games combined
with equal probability, therefore $\alpha = \frac{1}{2}$.
Setting ${\rm grad}[\langle X_c(t)
\rangle]$ equal to zero in Eq. \eqref{eqn:gradepsilon} gives:
\begin{multline}
\label{eqn:epsilon}
\e_{\rm{max}}
  =\frac{-4+p_1p_2+2p_3+2p_4-p_3p_4+p_2q_1+\ldots}
	{2(4+p_1+p_2-p_3-p_4+ \ldots}
	\\
   \frac{\ldots p_1q_2+q_1q_2+2q_3-p_4q_3+2q_4 -p_3q_4-q_3q_4}
        {\ldots q_1+q_2-q_3-q_4)}
	\\
\end{multline}
Appealing to the condition (Eq. \eqref{eqn:q1fair}) that $\{q_i\}$ is a fair
game for $\e=0$ and in the special case where $q_2=q_3$ and $p_2=p_3$, this
becomes:
\begin{multline}
\label{eqn:epsilon_spec_fair}
\e_{\rm{max}}
  =\frac{p_2[1+q_2(1+p_1-p_4)-q_4] + \ldots}
        {2(1-q_4) + \ldots}
	\\
   \frac{q_2[-3+p_4(2-q_2)+q_2+p_1q_2+q_4]}
        {2q_2(3+p_1-p_4)}
\end{multline}
\smallskip

\noindent Similarly in the general case, $\e_{\rm{max}}$ is given by:
\begin{multline}
\label{eqn:epsilon_gen_fair}
\e_{\rm{max}}
  =\frac{p_2 [1 - q_4 + q_2 p_1 + q_3 (q_4 - 1)] + \ldots}
        {2-2q_4 + \ldots}
	\\
   \frac{\ldots q_2[-3 + p_4 (2 - q_3) + q_3 + q_4] + \ldots }
        {\ldots 2q_2(4 + p_1 + p_2 - p_3 - p_4) + \ldots}
	\\
   \frac{\ldots q_2^2 p_1 + p_3 q_2 (2 - p_4 - q_4)}
        {\ldots 2 q_2 (q_2 - q_3 - q_4) + 2 q_3 (q_4 - 1)}
	\\
\end{multline}
\vfill


\end{document}